# Sign-resolved nanoscale readout and control of hidden antiferromagnetic spin order

A. Schmid[1]†, D. Siebenkotten[2,3]†, D. Dai[2]†, J. Godinho[1], T. Ostatnický[4], N. Zou[5], Y. Zhang[5,6], J. Železný[7], Z. Šobáň[7], F. Křížek[7], V. Novák[7], S. Fairman[2], A. Hoehl[2], A. Hertwig[8], T. Janda[1], M. A. Huber[1,9], R. Huber[1,9,10], B. Kästner[2], J. Wunderlich[1,7,10]*

*[1]Department of Physics, University of Regensburg, 93040 Regensburg, Germany.*
*[2]Physikalisch-Technische Bundesanstalt, 10587 Berlin, Germany.*
*[3]Institute of Physics (IA), RWTH Aachen University, D-52056 Aachen, Germany*
*[4]Faculty of Mathematics and Physics, Charles University, 121 16 Praha 2, Czech Republic.*
*[5]Department of Physics and Astronomy, University of Tennessee, Knoxville, TN 37996, USA*
*[6]Min H. Kao Department of Electrical Engineering and Computer Science, University of Tennessee, Knoxville, Tennessee 37996, USA*
*[7]Institute of Physics, Czech Academy of Sciences, Cukrovarnická 10, 160 00, Prague 6, Czech Republic*
*[8]Bundesanstalt für Materialforschung und -prüfung (BAM), Unter den Eichen 44 – 46, 12203 Berlin, Germany*
*[9]Center for Ultrafast Nanoscopy (RUN), University of Regensburg, 93040 Regensburg, Germany*
*[10]Halle-Berlin-Regensburg Cluster of Excellence CCE, University of Regensburg, 93040 Regensburg, Germany*
*†These authors contributed equally to this work.*
**Corresponding author. Email: joerg.wunderlich@ur.de*

Antiferromagnetic memories promise ultrafast, stray-field-free information storage. Yet perfect magnetic compensation conceals the information carrier itself: the sign of the Néel vector distinguishing two time-reversed states. Moreover, in future dense memories, the local polarity of Néel domains would need to be read out on the nanoscale. We make this hidden polarity visible in a fully-compensated, high-Néel-temperature, $\mathcal{PT}$-symmetric antiferromagnet by driving interband electric-dipole transitions by mid-infrared near-fields confined at a scanning probe. The excitation generates a Néel-order-dependent quantum-metric photocurrent, a Hall-like signal reversing with Néel order, which we term the optical nonlinear anomalous Hall effect. This optically induced electrical readout maps opposite Néel polarities with sub-100-nm resolution at room temperature and, combined with spin-orbit-torque writing, reveals Néel-texture polarization and reversible domain-wall motion, establishing electrical-write/optoelectronic-read antiferromagnetic functionality.

**One-sentence summary:** Near-field induced quantum-metric photocurrent nanoscopy reveals the hidden polarity of electrically written Néel order at the nanoscale.

Magnetic compensation is the source of both the technological promise and the readout problem of antiferromagnetic storage (*1–8*). By eliminating net magnetization, a fully compensated Néel state suppresses stray-field-mediated crosstalk and enables dense integration. The same absence of a net moment removes the linear dipolar coupling that makes the sign of a ferromagnetic bit directly visible. The stored binary variable is therefore the sign of the Néel order parameter, which distinguishes the two time-reversed states $+\boldsymbol{N}$ and $-\boldsymbol{N}$. In high-density devices, this variable would be encoded in nanoscale Néel domains, so readout must resolve not only the local orientation axis but also the polarity of each domain. This requirement is particularly relevant for uniaxial antiferromagnets, where the two time-reversed states are separated by a hard-axis anisotropy barrier that protects them against thermally driven reversal and makes them attractive nonvolatile binary states (*9–13*). Current pulses can write this variable through relativistic Néel spin-orbit torques (*3, 10, 13*), but reading its sign after writing requires a signal that is both local and odd under Néel vector reversal.

Existing antiferromagnetic readout techniques do not provide this combination of locality and sign sensitivity. In ferromagnets, moment reversal is detected directly through time-reversal-odd linear magnetotransport, magneto-optical, and thermomagnetic effects. In fully compensated conventional collinear antiferromagnets, however, analogous linear sign-sensitive responses are generally forbidden by magnetic symmetry. Synchrotron-based XMLD-PEEM is the established nanoscale benchmark for imaging antiferromagnetic domain axes and current-induced reorientation (*3, 14*), but its contrast is even in the Néel vector and therefore identical for $+\boldsymbol{N}$ and $-\boldsymbol{N}$. Related electrical and thermoelectric probes, including anisotropic-magnetoresistance-based readout and magneto-Seebeck microscopy (*15, 16*), can track domain reorientation or texture, but likewise do not provide time-reversal-odd contrast between 180°-reversed Néel polarities. Consequently, they cannot locate $+\boldsymbol{N}$ and $-\boldsymbol{N}$ domains.

Electrical nonlinear Hall physics provides a symmetry route to a macroscopic, current-path-averaged signal that is odd under reversal of the Néel order. In $\mathcal{PT}$-symmetric antiferromagnets, combined parity–time symmetry forbids Berry-curvature-driven linear anomalous Hall responses, while permitting second-order Hall responses whose sign reverses with the Néel vector (*17, 18, 19*).

Here, we translate this symmetry principle into a local optoelectronic photocurrent response in a fully compensated, high-Néel-temperature $\mathcal{PT}$-symmetric antiferromagnetic metal. Mid-infrared near-field excitation drives interband electric-dipole transitions that, through spin–orbit coupling, generate a Néel-order-dependent quantum-metric photocurrent, as theoretically predicted in Refs. *20, 21*. This local transverse photocurrent reverses with the Néel vector, constituting the first experimental observation of the optical nonlinear anomalous Hall effect (optical nl-AHE). Because its source is confined to the nanoscale near-field excitation volume, the readout directly maps $+\boldsymbol{N}$ and $-\boldsymbol{N}$ domains with sub-100-nm resolution at room temperature. Combined with spin-orbit-torque writing, optical nl-AHE nanoscopy distinguishes polarization of initially nanoscale textures from reversible displacement and reshaping of pre-existing domain walls, directly visualizing electrical writing and sign-resolved optoelectronic reading of hidden antiferromagnetic order.

**Near-field detection of Néel-odd photocurrents**

To search experimentally for the optical nl-AHE, we use mid-infrared near-field excitation in 20-nm-thick epitaxial films of the $\mathcal{PT}$-symmetric antiferromagnetic metal CuMnAs. The optical nl-AHE denotes a second-order, Hall-like photocurrent generated by the optical electric field and odd under Néel-vector reversal. We detect this local source current in an open-circuit geometry: charge redistribution creates a steady-state potential difference between remote contacts, which we define as the optical-nonlinear-AHE voltage, $V_{\mathrm{oNA}}$. Materials and methods are provided in the supplementary materials. All measurements were performed at room temperature in this high-Néel-temperature metallic antiferromagnet.

Fig. 1A sketches the experimental idea. In the s-SNOM experiment, a metallic atomic-force microscopy tip illuminated by mid-infrared light ($\lambda = 10$ μm) concentrates the electric field at the metallic CuMnAs surface into a strongly enhanced, sub-100-nm near field that is dominated by its surface-normal component and is approximately rotationally symmetric about the tip axis (supplementary materials). Based on second-order optical-response theory for $\mathcal{PT}$-symmetric antiferromagnets (*20–25, 34–37*), this local field is expected to drive interband electric-dipole transitions and inject a Néel-order-odd photocurrent; for the high-symmetry states relevant here ($\boldsymbol{N}$ parallel to $\langle 110\rangle$, and likewise $\langle 100\rangle$), the net current should be perpendicular to $\boldsymbol{N}$. The central experimental question is therefore whether this symmetry-allowed, Néel-order-odd current can be converted into a measurable $V_{\mathrm{oNA}}$ and, once verified, used as the contrast mechanism for high-resolution, sign-resolved antiferromagnetic nanoscopy.

CuMnAs provides the required symmetry (Fig. 1B). In the non-magnetic phase, the crystal is centrosymmetric; antiferromagnetic order breaks inversion and time-reversal symmetries separately while preserving combined $\mathcal{PT}$-symmetry. A second-order conductivity tensor that is odd in the Néel order is therefore allowed only in the antiferromagnetic phase; reversing the Néel vector should reverse the local optical nl-AHE current (*20–26*). To test this, we scan the illuminated tip across a 1 μm-wide CuMnAs bar and record $V_{\mathrm{oNA}}$ between the outer contacts (Fig. 1C). If antiferromagnetic domains are larger than the near-field excitation area, the locally injected photocurrent should generate a voltage whose sign follows the local Néel polarity. Remarkably, the resulting $V_{\mathrm{oNA}}$ map indeed shows submicron regions of opposite sign, providing the first experimental indication that the anticipated optical nl-AHE exists. Next, we will establish its symmetry and microscopic origin.

**Electrical writing and Néel order reversal**

The large as-grown domains visible in this narrow bar are attributed to strain-driven variations of magnetic anisotropy caused by lattice-mismatch relaxation, rather than to dipolar interactions as in ferromagnets. Similar strain-induced effects have been reported in patterned thin films with negligible dipolar coupling, including low-moment diluted magnetic semiconductors (*27*) and antiferromagnets (*28*). Their tendency to form large domains weakens with increasing device width, motivating the wider-device experiments below, in which current pulses prepare defined Néel-vector orientations and verify the symmetry of the anticipated optical nl-AHE. In the relaxed ground state, the unstructured film forms nanoscale domains with Néel vectors aligned along different crystallographic axes (*29, 30*). We first use a large, 5 µm-wide, four-terminal crossbar (Fig. 2A) where patterning does not substantially affect the domain structure of the unstructured film and which enables electrical writing and photocurrent readout.

For the crossbar, we define $V_{\mathrm{oNA}} = (V_{A^+} - V_{A^-}) - (V_{B^+} - V_{B^-})$ as the difference of the simultaneously recorded potential drops along the A and B axes, corresponding to the net potential drop along the upper-right to lower-left diagonal (Fig. 2A; supplementary materials). Height modulation of the s-SNOM tip isolates the near-field contribution to the recorded AC voltage.

Because the domain size is comparable to the near-field excitation area ($\sim$20 nm), the associated local photocurrents are expected to average out, yielding no resolvable domain contrast. Consistent with this, the initial $V_{\mathrm{oNA}}$ map (Fig. 2B) shows only a weak residual background, indicating that the nanoscale domains share a slight common component perpendicular to the upper-right to lower-left diagonal.

To overwrite the nanoscale texture and create macroscopic areas with a defined Néel vector, we apply 10-ms-long current pulses of approximately 40 mA along the upper-right to lower-left diagonal by shorting the contact pairs ($V_{A^+}, V_{B^-}$) and ($V_{A^-}, V_{B^+}$). Simulations show that the current density, and hence the staggered spin–orbit field, peaks at the two inner corners transverse to the pulse direction (fig. S2A). Above the switching threshold, the Néel vector reorients perpendicular to the current (*3, 10, 17*), producing two corners with parallel average

Néel vectors. Reversing the pulse polarity reverses the Néel vector in the same regions, yielding a 180°-switched configuration. After each pulse, we scan the illuminated s-SNOM tip across the device and record $V_{\mathrm{oNA}}$ using the same contact configuration. Fig. 2C shows strong signals at the two pulse-affected corners, indicating that the optical nl-AHE current in these written regions is directed along the writing current.

Because spin–orbit-torque switching aligns the Néel vector perpendicular to the pulse current, this is consistent with a transverse Hall-like response expected for the high-symmetry $\langle 110 \rangle$ states in our rotationally symmetric near-field geometry. Reversing the pulse polarity flips the oNA-voltage sign at both corners (Fig. 2D), evidencing that the photocurrent is locked to the Néel-vector orientation and changes sign under time reversal.

Orthogonal-channel measurements resolve no longitudinal photocurrent contribution from the pulse-affected corners (supplementary materials; figs. S3 and S4). Together with the sign reversal under Néel-vector inversion, this establishes a time-reversal-odd Hall-like photocurrent for the pulse-written states (*21*). Finally, the $V_{\mathrm{oNA}}$ signals localized at the two pulse-affected corners disappear after approximately 15 hours, consistent with relaxation towards the initial sub-resolution multidomain state (*30*) (supplementary materials; fig. S3C and D).

**Microscopic origin of the optical nonlinear anomalous Hall effect**

Having established experimentally that the optical nl-AHE is locked to the Néel vector, we now address its microscopic origin. The relevant response is a time-reversal-odd interband injection current allowed by the broken inversion and time-reversal symmetries of the antiferromagnetic state (*20, 21, 23–25*). We evaluate it from a second-order Kubo formalism in the clean limit and combine it with the simulated s-SNOM near-field distribution.

In general, the injection current can be written as $j_a = \sigma_{abc}^{(2)}(\omega)[E_b(\omega)E_c^*(\omega) + E_b^*(\omega)E_c(\omega)]$, where the nonlinear conductivity tensor $\sigma_{abc}^{(2)}$ is obtained from a second-order Kubo formalism (*20*):

$$\sigma_{abc}^{(2)}(\omega) = \tau \frac{\pi e^3}{\hbar^2} \sum_{n \neq m} \int_{\mathrm{BZ}} \frac{d^3k}{(2\pi)^3} \, \Delta v_a^{nm}(\mathbf{k}) r_b^{nm}(\mathbf{k}) r_c^{mn}(\mathbf{k}) f_{nm}(\mathbf{k}) \delta(\varepsilon_m(\mathbf{k}) - \varepsilon_n(\mathbf{k}) - \hbar\omega). \quad (1)$$

Equation (1) describes the generation of photoexcited electron–hole pairs whose group-velocity contributions do not cancel when the electronic structure lacks the symmetries that enforce $\mathbf{k} \leftrightarrow -\mathbf{k}$ compensation. Here $r_b^{nm}$ is the interband Berry connection between bands $n$ and $m$, $\Delta v_a^{nm} = v_a^{mm} - v_a^{nn}$ is the corresponding group-velocity difference, $f_{nm} = f_n - f_m$ enforces occupied-to-empty transitions, with $f_i$ the Fermi distribution function of band $i$;

and $\tau$ is the lifetime of the optically excited carriers (assumed equal for all excited states). In $\sigma_{abc}^{(2)}$, the index $a$ denotes the injected current direction, while $b$ and $c$ label Cartesian components of the driving electric light field.

In our near-field geometry, the incident laser beam is linearly polarized, and its coupling to the metallic tip generates a strongly enhanced local field whose dominant component is oriented along the surface normal, $E_z$. Finite in-plane components, $E_x$ and $E_y$ are also present and are included in our simulations. In addition, the in-plane near-field components are rotationally symmetric about the s-SNOM tip position (Fig. 3B, inset, and supplementary materials), such that the spatially integrated photocurrents driven by the mixed field products $E_x \cdot E_y$, $E_z \cdot E_x$ and $E_z \cdot E_y$ cancel in our near-field geometry. Moreover, the corresponding second-order conductivity tensor elements associated with these mixed field components are negligible (supplementary materials, fig. S5). The measured net response is thus dominated by contributions from the field products $E_x \cdot E_x$, $E_y \cdot E_y$, and $E_z \cdot E_z$ only. Accordingly, for the calculated injection current and corresponding $V_{\mathrm{oNA}}$ map shown in Fig. 3B and D, respectively, we retain the full tensor structure and scale the tensor elements by the corresponding simulated near-field components. Throughout, we use Cartesian axes defined by $x \parallel [110]$ and $y \parallel [1\bar{1}0]$.

In the above calculations, we neglect disorder-related scattering contributions such as skew scattering and side-jump processes, which can also contribute to the optical nonlinear Hall responses. Disorder effects in $\mathcal{PT}$-symmetric antiferromagnets, including skew-scattering-related nonlinear Hall and chiral photocurrent mechanisms, have recently been analyzed theoretically (*31, 32*), but their quantitative weight under mid-infrared interband excitation in our near-field geometry remains to be established. We therefore focus on the clean-limit injection current of Eq. (1), whose magnitude scales linearly with the optically excited carrier lifetime $\tau$.

To connect this microscopic picture with the experiment, we performed first-principles density functional theory calculations (*33*), focusing on configurations with the Néel vector aligned along $[110]$, as in the switching experiments shown in Fig. 2. For $\boldsymbol{k} \parallel \boldsymbol{N}$, the calculated dispersion remains symmetric under $\boldsymbol{k} \to -\boldsymbol{k}$, whereas for $\boldsymbol{k} \perp \boldsymbol{N}$ it exhibits a pronounced $+\boldsymbol{k}$ versus $-\boldsymbol{k}$ asymmetry (Fig. 3A). This asymmetry yields a net injected photocurrent whose time-reversal-odd part is dominated by the component transverse to $\boldsymbol{N}$. Consistently, the computed injection-current conductivity tensor elements (Fig. 3B) show finite time-reversal-odd components only for current directions perpendicular to $\boldsymbol{N}$, while the corresponding components parallel to $\boldsymbol{N}$ vanish within numerical accuracy. In combination with the rotationally symmetric near-field geometry discussed above, this explains the experimentally observed transverse net response without implying that the same relation must hold for an arbitrary in-plane Néel vector direction.

Using the pulse-written switched Néel-order texture, the calculated second-order nonlinear conductivity tensor and the s-SNOM tip–induced near-field distribution inside the film, we compute mid-infrared $V_{\mathrm{oNA}}$ maps for the crossbar device (supplementary materials). Because current-induced alignment of the Néel vector parallel to the staggered spin–orbit field occurs only where the local current density exceeds the switching threshold, namely, at the pair of inner corners transverse to the pulse direction (Fig. 3C), the simulated corner signal (Fig. 3D) (*34, 35*) reproduces the spatial localization of the measured $V_{\mathrm{oNA}}$ map (Fig. 2C). Quantitative agreement with the experimental signal magnitude requires an effective relaxation time $\tau = 235\ \mathrm{fs}$ used to calculate the microscopic conductivity in the clean limit. The $\tau$ should be interpreted as a phenomenological, mid-infrared-averaged interband dephasing time, rather than as the equilibrium Drude transport scattering time of a metal or semimetal such as CuMnAs.

**Sign-resolved tracking of domain-wall motion**

To exploit the optical nl-AHE readout in devices with a pre-existing magnetic texture, we repeated the write–read protocol on 1-μm-wide crossbars patterned from the same 20 $nm$-thick CuMnAs film used in Figs. 1 and 2. In these narrow devices, the pristine state already exhibits extended antiferromagnetic domains, consistent with strain-relaxation-induced variations of the magnetic anisotropy. Diagonal writing pulses reversibly reconfigure this texture, with the largest changes occurring in the central overlap region and at the adjacent inner corners, where the pulse current density is highest. Fig. 4 focuses on the current-affected regions and on the readout channels most sensitive to the corresponding writing configurations. The diagonal voltage drops are obtained from the simultaneously recorded contact-pair voltages $V_A = V_{A^+} - V_{A^-}$ and $V_B = V_{B^+} - V_{B^-}$. We then analyze the linear combinations $V_A + V_B$ and $V_A - V_B$, which correspond to the two orthogonal diagonal readout axes. Pulse-polarity-dependent contrast is strongest at the inner corners affected by the writing current. For a given writing pulse, the largest changes occur in the channel that measures the potential difference along the diagonal writing axis; when the writing pulse is rotated by 90°, the dominant response switches to the orthogonal channel. This behavior is consistent with the Hall-like photocurrent orientation identified above for pulse-written high-symmetry $\langle 110 \rangle$ Néel states (Figs. 2 and 3).

In contrast to the 5 $\mu m$-wide device in Fig. 2, where pulses overwrite an initially sub-resolution texture, the 1-μm-wide crossbars are reconfigured mainly by displacement and reshaping of pre-existing extended domains. Domain-wall motion and defect pinning can therefore produce a spatially complex response (*36, 37*), so the local $V_{\mathrm{oNA}}$ polarity does not everywhere follow the writing-current direction. Nevertheless, repeated maps reveal reversible displacement of sign-contrasted domain features upon pulse-polarity reversal, providing direct evidence for electrically driven antiferromagnetic domain-wall motion tracked locally with nanoscale resolution. Because the $V_{\mathrm{oNA}}$ contrast retains its sign, this polarity-resolved tracking goes beyond both spatially averaged switching measurements and high-resolution XMLD-based antiferromagnetic microscopies, whose contrast is even under time reversal. The complete, uncropped dataset for both readout combinations over the full crossbar is provided in fig. S6.

Finally, we benchmark our optical nl-AHE imaging method on 50 $nm$-thick CuMnAs/GaP devices, whose domain pattern is governed by microtwin/defect-line pinning (*29*). The optical nl-AHE maps reproduce the stripe-like defect-correlated morphology known from XMLD-PEEM, while additionally resolving time-reversal-odd contrast between 180°-reversed domains (supplementary materials; fig. S7).

**Implications for hidden Néel order readout**

In summary, we report the experimental observation of a long-sought-after light–spin interaction mechanism in the $\mathcal{PT}$-symmetric antiferromagnet CuMnAs. Using mid-infrared near-field excitation, the optical nonlinear anomalous Hall effect generates a local transverse photocurrent that is odd under Néel reversal and therefore enables nanoscale imaging of the Néel texture with direct sensitivity to 180°-reversed domains. Our new sign-sensitive nanoscopy reveals reversible current-pulse-induced writing of the Néel vector using spin-orbit torque in two write modes: the polarization of initially nanometre-scale domains into larger spin-orbit-field-aligned regions, and domain-wall motion in pre-existing extended domains. Together with first-principles calculations and device-level modeling, these measurements show that, for the symmetry-selected in-plane states probed here, the measured $V_{\mathrm{oNA}}$ signal is consistent with a Néel-order-odd photocurrent directed perpendicular to the local Néel vector. The optical nl-AHE places this readout in the broader context of nonlinear Hall physics, while establishing a distinct capability. In $\mathcal{PT}$-symmetric antiferromagnets, combined parity–time symmetry suppresses Berry-curvature-driven linear Hall responses, but permits second-order nonlinear Hall currents that are odd in the Néel order through quantum-metric response tensors, such as the quantum metric and Berry-connection polarizability. Electrical nonlinear anomalous Hall measurements (*17, 49*), including our earlier work (*17*), have shown that 180° Néel-vector reversal can be converted into a time-reversal-odd transverse transport response. In conventional device geometries, however, the driving field and detected voltage extend over the current path; the signal is therefore spatially averaged and cannot locate $+\boldsymbol{N}$ and $-\boldsymbol{N}$ domains. By contrast, the optical nl-AHE is generated by interband electric-dipole transitions within the nanoscale near-field volume of a scanning probe, transforming nonlinear Hall sign sensitivity into a local photoelectric source. This route is also distinct from compensated altermagnets, where symmetry can make the leading time-reversal-odd nonlinear response third order in the electric field, even though time-reversal-odd linear responses may be allowed when magnetic symmetry permits momentum-dependent spin polarization (*38–43*).

Optical nl-AHE thus offers tabletop, laboratory-scale antiferromagnetic nanoscopy complementary to synchrotron-based XMLD microscopy, while overcoming its fundamental insensitivity to the sign of $\pm\boldsymbol{N}$. More generally, our results establish nonlinear Hall photocurrents as a route to read out hidden order in $\mathcal{PT}$-symmetric antiferromagnets and motivate future ultrafast optoelectronic read/write concepts of antiferromagnetic order. The room-temperature operation in a high-Néel-temperature metallic $\mathcal{PT}$-symmetric antiferromagnet is essential for the technological relevance of this read-write principle.

During the preparation and review of this manuscript, we became aware of independent studies carried out in parallel by Dong *et al.* (*44*) and Tian *et al.* (*45*), which also report quantum-metric or photocurrents in the low-Néel-temperature van der Waals antiferromagnets CrSBr and $MnPSe_3$, respectively. Our work extends these findings by demonstrating that a Hall-like quantum-metric photocurrent can serve as a local nanoscale probe in metallic antiferromagnet with Néel temperatures far above room temperature. Our near-field approach enables nanoscale imaging of Néel-order polarity, local readout of electrically written spin–orbit-torque states, and direct visualization of current-driven domain-wall motion.

**Acknowledgments:**
J.W. thanks Leonid E. Golub and Yuriy Mokrousov for helpful discussions.

**Funding:** This work was supported by the Deutsche Forschungsgemeinschaft (DFG, German Research Foundation) through SFB 1277 (project ID 314695032), GRK 2905 (project ID 502572516), project no. 529998081, and grants INST 89/562, WU 883/4, HU 1598/8 and KA 2866/2; and through the Germany's Excellence Strategy – EXC3112/1 – 533767171 (Center for Chiral Electronics).

Additional support was provided by the Dioscuri Program LV23025 funded by the Max Planck Society (MPG) and the Ministry of Education, Youth and Sports of the Czech Republic (MEYS) through e-INFRA CZ (ID:90254); by the Czech Science Foundation (GAČR) through projects 21-28876J and 25-18244S; by the Czech Academy of Sciences through the Lumina Quaeruntur fellowship LQ100102602, by TERAFIT (CZ.02.01.01/00/22_008/0004594), co-financed by the European Union's Horizon 2020 research and innovation programme under the Marie Skłodowska-Curie grant agreement no. 861300 (COMRAD).

N.Z. and Y.Z. were supported by the Max Planck Partner Lab for Quantum Materials.

**Author contributions:**

Performance of experiments and analyzation of data: A.S., D.S., D.D., J.G., A.Ho., A.He., T.J., M.A.H., R.H., J.W., B.K.

First-principles density functional theory calculations: N.Z., J. Ž., Y.Z.

Numerical simulations: D.S., D.D., S.F., T.O.

Sample growth and characterization of antiferromagnetic thin films: F.K., V.N.

Fabrication of Cross-bar devices: Z.Š.

Writing of manuscript: D.D., A.S., D.S., J.G., T.O., N.Z., M.A.H., R.H., B.K., J.W.

Discussion of results: all authors

Design of experiments: B.K.

Idea and plan of experiments: J.W.

**Competing interests:**

The authors declare no competing interests.

**Data, code, and materials availability:**

All data needed to evaluate the conclusions in the paper are present in the paper and/or the supplementary materials. Source data and analysis code are available at [repository/DOI to be inserted before publication].

**Supplementary Materials**

- Materials and Methods
- Supplementary Text
- Figs. S1 to S7
- References (46-51)

## Figures

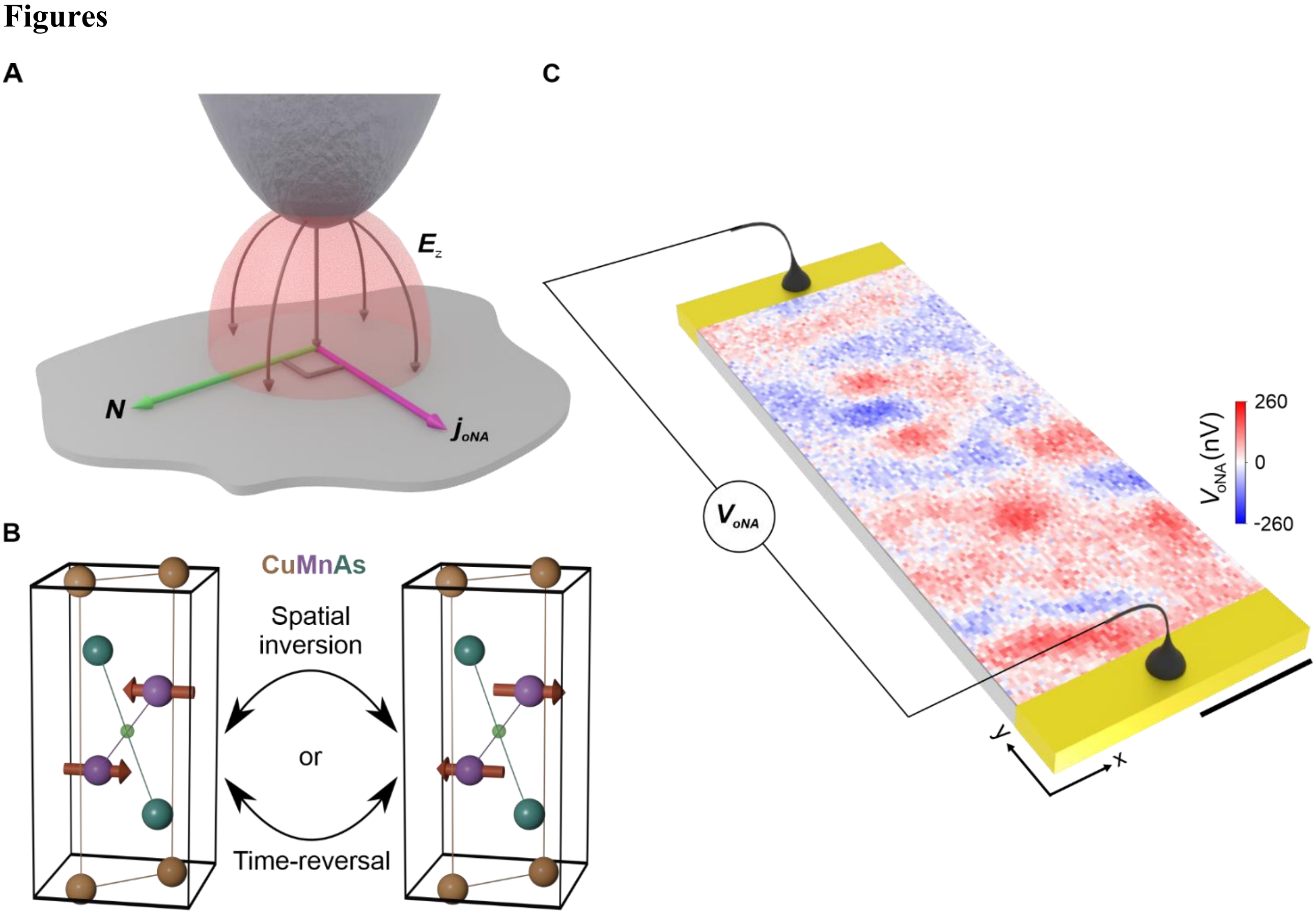

**Fig. 1. Near-field photocurrent detection of the optical nl-AHE in CuMnAs.** (**A**) Schematic of near-field excitation beneath a metallic AFM tip. Incident mid-infrared light ($\lambda = 10$ μm) generates a nanoscale near field dominated by the out-of-plane component $E_z(t)$ (red hemisphere), driving interband dipole transitions within the film. The resulting photocurrent $j_{\mathrm{oNA}}$ (purple arrow) is shown transverse to the Néel vector $\boldsymbol{N}$ (green arrow) for the high-symmetry in-plane states relevant to the experiment, consistent with the optical nl-AHE symmetry. (**B**) Schematic representation of the CuMnAs crystal structure. In the non-magnetic state (ignoring the brown arrows indicating the antiferromagnetically coupled magnetic moments of the Mn atoms), CuMnAs is centrosymmetric (green sphere highlights the inversion center); in the antiferromagnetic state, spatial inversion $P$ and time-reversal $T$ symmetries are broken individually, but CuMnAs preserves combined $PT$ symmetry. (**C**) Color-coded oNA-voltage ($V_{\mathrm{oNA}}$) map measured between the two contacts (yellow) while scanning the irradiated tip across a 1-μm-wide CuMnAs bar. Red and blue contrasts correspond to opposite voltage polarities and reveal individual antiferromagnetic domains (scale bar, 500 nm).

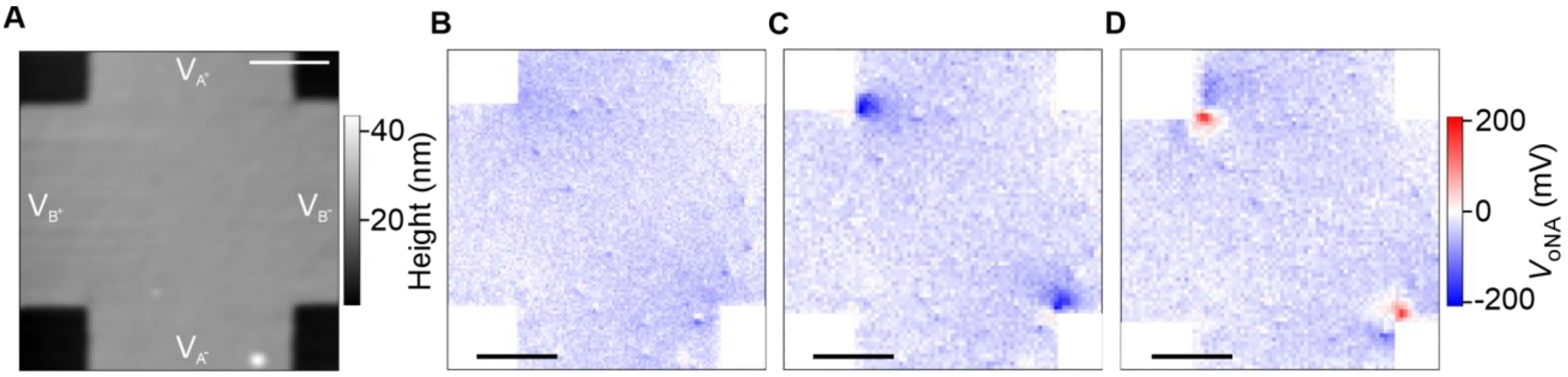

**Fig. 2. Current-induced control of the Néel vector and detection of the optical nl-AHE.** (**A**) AFM topography of the 5 μm-wide CuMnAs cross, acquired simultaneously with the oNA-voltage using an illuminated s-SNOM tip. (**B**) Initial oNA-voltage map $\Delta V = (V_{A^+} - V_{A^-}) - (V_{B^+} - V_{B^-})$, showing little contrast due to spatial averaging (white/black scale bars, 2 μm). (**C**) The first written-state oNA-voltage map after pulsing, showing corner-localized signals consistent with the Hall-like current orientation expected for the pulse-written high-symmetry states along $\langle 110 \rangle$. (**D**) oNA-voltage map after the reversed-polarity pulse, showing sign reversal of the optical nl-AHE signal and confirming Néel-vector reversal at both corners (time-reversal-odd response).

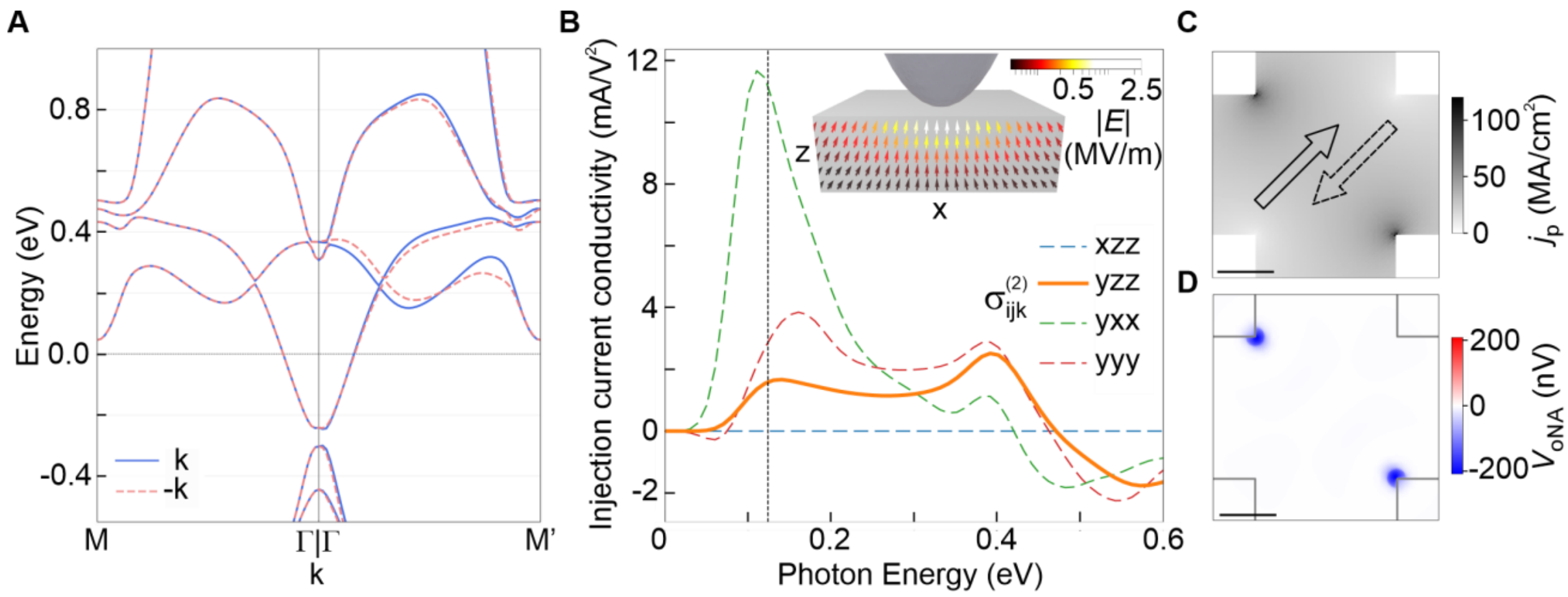

**Fig. 3. Microscopic origin and quantitative modeling of the optical nonlinear anomalous Hall effect.** (**A**) First-principles band dispersion of CuMnAs for $\boldsymbol{N}$ parallel to [110], corresponding to the $M$–$\Gamma$ direction, shown for opposite crystal momenta $+\boldsymbol{k}$ (solid) and $-\boldsymbol{k}$ (dashed) along the orthogonal high-symmetry directions $M$–$\Gamma$ (parallel to $\boldsymbol{N}$) and $\Gamma$–$M'$ (perpendicular to $\boldsymbol{N}$). A pronounced $+\boldsymbol{k}$ versus $-\boldsymbol{k}$ asymmetry appears for momenta orthogonal to $\boldsymbol{N}$, yielding a time-reversal-odd interband injection current whose transverse component dominates for this high-symmetry in-plane orientation. (**B**) Calculated injection-current conductivity tensor elements $\sigma^{(2)}_{abb}(\omega)$ that contribute to the injection current in the s-SNOM geometry, evaluated using a common phenomenological broadening $\hbar/\tau \approx 3$ meV, corresponding to $\tau = 235$ fs. Tensor components that generate a current perpendicular to the Néel vector are finite, whereas the corresponding component parallel to the Néel vector vanishes within numerical accuracy for this orientation. The vertical dotted line marks the photon energy of the mid-infrared excitation ($\lambda = 10$ µm). The bold orange curve denotes $\sigma^{(2)}_{yzz}$, the dominant contribution because $|E_z|$ is largest (see inset). The inset shows the simulated electric-field distribution inside the 20-nm-thick CuMnAs layer for a tip positioned 1.35 nm above the surface and irradiated with 15 mW incident power (see supplementary materials; scale bar, 10 nm). (**C**) Simulated current-density distribution in the 5-µm-wide symmetric crossbar patterned from the 20-nm-thick CuMnAs film for a 40-mA diagonal writing pulse applied from the upper right to the lower left (or vice versa). The current density is enhanced, and the switching threshold is exceeded at the pair of inner corners transverse to the pulse direction (see supplementary materials and fig. S2A). (**D**) Simulated oNA-voltage map obtained by combining the calculated nonlinear conductivity tensor with the FEM near-field distribution and the pulse-written Néel-order texture, modeled as uniformly switched inner-corner regions. The simulated map reproduces the spatial localization and magnitude of the measured oNA-voltage signal in Fig. 2C (scale bars, 2 µm).

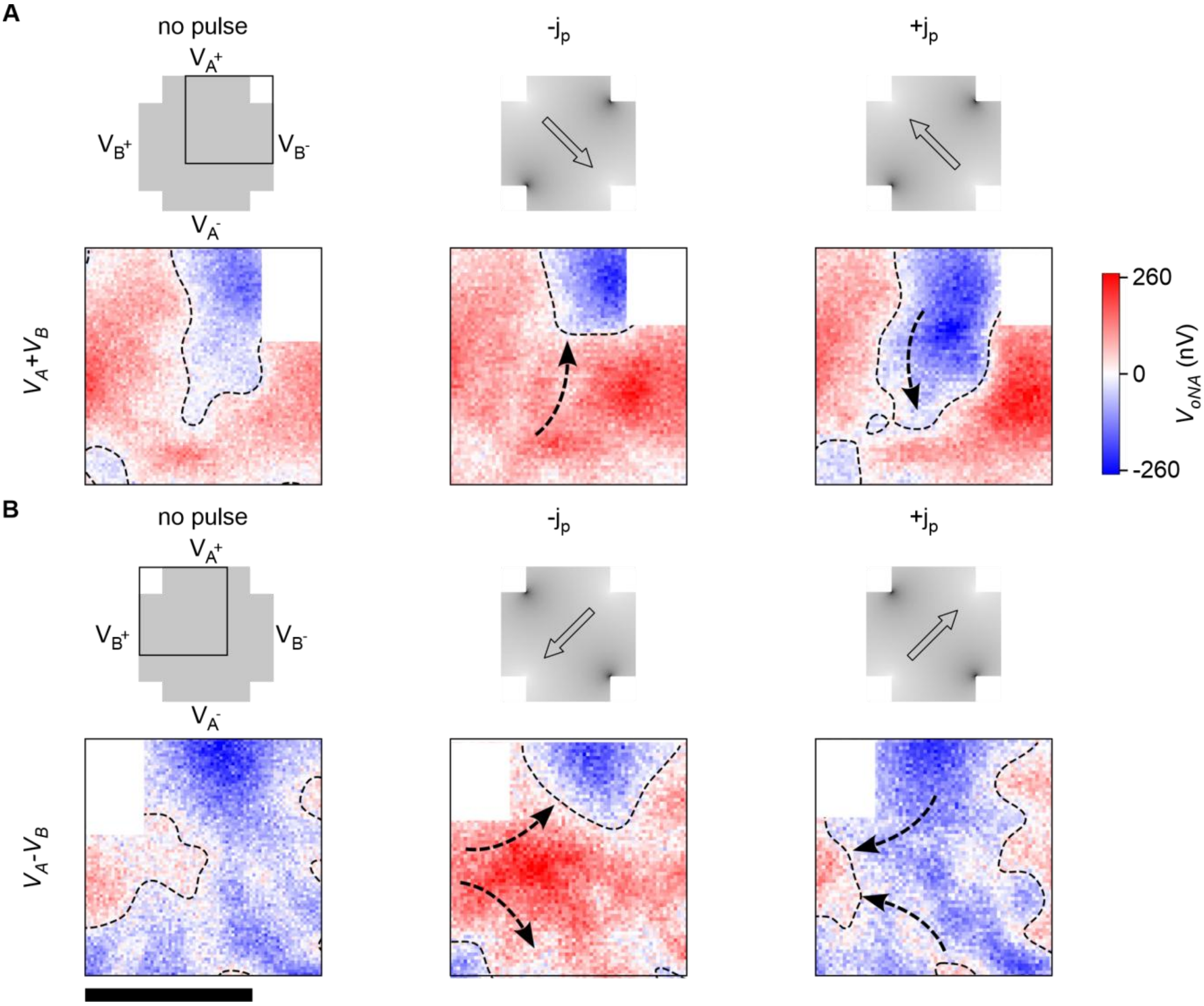

**Fig. 4. Optical nl-AHE imaging of reversible Néel-texture reconfiguration in a 1 $\mu m$-wide CuMnAs crossbar.** Cropped oNA-voltage maps of selected current-affected regions of a 20 $nm$-thick CuMnAs crossbar with a pre-existing antiferromagnetic domain texture. The contact-pair voltages are defined as $V_A = V_A^+ - V_A^-$ and $V_B = V_B^+ - V_B^-$. The no-pulse schematics define the voltage contacts and the displayed field of view; the grayscale maps in the pulsed columns show the simulated magnitudes of current-density distributions with arrows indicating the average current direction. (**A**) oNA-voltage maps of the upper-right corner in the $V_A + V_B$ readout channel. (**B**) oNA-voltage maps of the upper-left corner in the $V_A - V_B$ readout channel. In each row, the left map shows the pre-pulse texture, whereas the middle and right maps show the same region after writing pulses of opposite polarity along the corresponding diagonal. Dashed contours mark selected domain boundaries, and curved arrows highlight the reversible displacement and reshaping of domain features induced by the pulses. The largest polarity-dependent changes occur near the inner corners, where the simulated current density is highest. A common color scale is used for all oNA-voltage maps (scale bar, 500 nm). The complete, uncropped dataset for both readout combinations over the full crossbar is provided in fig. S6.

# Supplementary Materials for

## Sign-resolved nanoscale readout and control of hidden antiferromagnetic spin order

A. Schmid, D. Siebenkotten, D. Dai, J. Godinho, T. Ostatnický, N. Zou, Y. Zhang5, J. Železný, Z. Šobáň, F. Křížek, V. Novák, S. Fairman, A. Hoehl, A. Hertwig, T. Janda, M. A. Huber, R. Huber, B. Kästner, J. Wunderlich†

†Corresponding author: Email: joerg.wunderlich@ur.de

**The PDF file includes:**

Materials and Methods
Supplementary Text
Figs. S1 to S7
References (46-51)

## Materials and Methods

Device fabrication

The CuMnAs thin film was grown on (001) GaP substrate by molecular beam epitaxy. First, the native oxide was desorbed at 640 °C (measured optically by absorption edge) under P2 overpressure. A 50 nm thick GaP buffer was grown at 550 °C at growth rate of 1.13 nm/min. Subsequently, the sample was cooled down to 220 °C and a 20 nm thick CuMnAs layer was grown (0.35 nm/min) with fluxes calibrated to form 1:1:1 stoichiometric films. We note that the As4 flux is fine-tuned to obtain the ideal stoichiometry, which then results to formation of films with conformal morphology and reduced defect density (in comparison to Ref. *46*). After CuMnAs growth, the sample was cooled to around −20 °C and a 3 nm thin film of Al was deposited. The purpose is to prevent oxidation of the CuMnAs film at ambient conditions after natural formation of stable AlOx.

The wafers were then patterned into Hall bar and cross devices by electron beam lithography with a positive A4 PMMA resist. The structures are defined by a wet chemical process, where the etchant is based on the diluted $H_2PO_4$ acid. The electrical contacts were fabricated via a lift-off process using Cr/Au metallization with thicknesses of 4 nm and 80 nm, respectively.

Microscopic theory of the optical nonlinear anomalous Hall effect

Density functional theory calculations were performed using the full-potential local-orbital (FPLO) code (*33*). The exchange-correlation potential was parameterized within the generalized gradient approximation (GGA) using the Perdew-Burke-Ernzerhof (PBE) functional (*47*). The calculations were carried out in a fully relativistic mode, solving the four-component Kohn-Sham-Dirac equation to account for spin-orbit coupling. The integration over the Brillouin zone was performed using the tetrahedron method with a 12×12×12 k-mesh grid. A spin-polarized setup was employed, initialized with magnetic moments on the Manganese (Mn) sites with orientation along [110] direction.

To calculate the magnetic injection current on a finer k-grid, we further employed Wannier functions to construct a tight-binding model. The projection window included the energy range from −13.45 eV to 0 eV. The basis set for the projection consisted of Cu 4s, 3d; Mn 4s, 3d; and As 4s, 4p orbitals. Finally, the magnetic injection current was calculated using the HopTB package (*20, 48*). In these calculations, we adopted a dense k-grid of 200×200×200 and a Gaussian broadening of 3 meV.

Experimental Setup

All measurements were performed with commercially available s-SNOM setups (neasnom, attocube systems GmbH), depicted schematically in Fig. S1. Mid-infrared laser light of $\sim 10\ \mu\mathrm{m}$ wavelength from quantum cascade lasers (MIRcat, DRS Daylight Solutions) is focused on PtIr-coated atomic force microscopy tips (25PtIr200B-H, Rocky Mountain Nanotechnology LLC), where near-fields are excited at the tip apex, which excite local photocurrents as described in the main text. The resulting voltage is simultaneously measured at both pairs of remote contacts along the sample cross' axes and amplified by SR560 low noise amplifiers (Stanford Research Systems) at a factor of $5 \times 10^4$, along with the sample topography. Due to the small extent of the near-fields compared to the large focal spot size, s-SNOM is plagued by significant background contributions. To suppress this background, the atomic force microscope is operated in tapping mode, i.e., the tip is near-resonantly oscillated above the sample at the scale of the near-field decay (*49*). The photocurrent signal is then analysed at higher harmonics of the tapping frequency. At these harmonics, spatially highly non-linear near-fields contribute significantly more than the far-field background, which is essentially linear at the scale of the oscillation amplitude. In the mid-infrared, the background is generally suppressed sufficiently at the second harmonic, which we also confirmed in our photocurrent simulations.

Calculation of the local electric near-field in the SNOM experiment

Finite-element method (FEM) simulations of the tip-induced near-field were performed using JCMwave (*34, 35*). The model system comprises a $5\ \mu\mathrm{m}$ long metallic AFM tip above a CuMnAs/GaP heterostructure consisting of a $20\ nm$ CuMnAs film on a $980\ nm$-thick GaP substrate. The system is illuminated by vertically polarized mid-infrared radiation ($\lambda = 10\ \mu\mathrm{m}$) incident at 30° with respect to the sample plane. The permittivity of the CuMnAs layer at this wavelength was determined by a combination of ellipsometry and transport measurements as $\varepsilon_{\parallel} = -197 + 295i$ or the in-plane component and $\varepsilon_{\perp} = -39 + 59i$ for the perpendicular to plane component.

In the experiment, the tip oscillates in tapping mode, resulting in a periodically varying tip–surface distance. To account for this, the simulation was repeated for several tip heights. The simulation was performed in 3D using a rotational-symmetry approach, providing access to all three components of the electric field.

The Inset in Fig. 3B shows a representative distribution of the optical electric field in the x–z plane ($y = 0$) for a metallic AFM tip positioned 1.35 nm above the CuMnAs/GaP sample.

Arrows indicate the local field direction within the 20 nm thick CuMnAs film; the colour encodes the field magnitude (arbitrary units, normalised as in the colour bar). The incident field was calculated assuming diffraction limited focusing of 15 mW incident power. Within the simulation we assume plane wave excitation. The field distribution inside our CuMnAs layer causes the locally generated photocurrent via the nonlinear conductivity tensor. The relevant region is shown in the inset in Fig. 3B as the gray layer in the image. The near field is strongly enhanced and confined to the immediate vicinity of the surface beneath the tip apex (red arrows). Importantly, appreciable in-plane field components are also present and are therefore included in the photocurrent calculation, as discussed in Supplementary Materials.

## Supplementary Text

Numerical simulations of experimentally detected nonlinear optical anomalous Hall voltage

We performed numerical simulations of the optical nonlinear anomalous Hall effect (optical nl-AHE) in order to support our interpretation of the effect in terms of a locally generated nonlinear photocurrent that is converted into a detectable open-circuit voltage. The simulations comprise two steps: (i) modeling the pulse-written antiferromagentic domains texture created by current pulse, and (ii) modeling the near-field readout of this texture under local mid-infrared excitation at the apex of a metallic s-SNOM tip. We considered the full three-dimensional geometry of the crossbar-shaped 20 nm thick CuMnAs sample with 5 μm wide and 20 μm long arms.

Domain formation by a current pulse

To model the pulse-written domains, we first compute the pulse current distribution in the crossbar by applying electrostatic boundary conditions at the terminals.

The electrostatic potential is set to zero at the terminal edges labelled $V_{A+}$ and $V_{B-}$ in Fig. 2 in the main text, while the potentials at $V_{A-}$ and $V_{B+}$ are set to values chosen such that the total net current through the device is 40 mA in the sample. We then define a threshold current density $j_{th} = 4.2 \times 10^7\ \mathrm{A\,cm^{-2}}$, above which the Néel vector is assumed to reorient perpendicular to the local pulse current. Regions where $j(\mathbf{r}) > j_{th}$ are treated as compact, uniformly switched domains with the Néel vector along the diagonal direction (as in the inset of Fig. S2A), whereas the remaining parts of the device are assumed to consist of small, randomly oriented fragmented domains that do not yield a net near-field signal.

Conversion of the source photocurrent into a voltage signal

The source photocurrent $\boldsymbol{j}_{oNA}$ drives a redistribution of charge until a steady-state electrostatic potential $\varphi(\boldsymbol{r})$ builds up, producing an Ohmic return current that enforces the open-circuit condition. We write the total current density as $\boldsymbol{j}_{tot} = \boldsymbol{j}_{oNA} - \sigma^{(1)}\nabla\varphi$, where $\sigma^{(1)}$ is the sample linear (ohmic) conductivity of the CuMnAs film. Charge conservation $\nabla \cdot \boldsymbol{j}_{tot} = 0$, together with the Neumann boundary conditions (vanishing normal component of $\mathbf{j}_{\mathrm{tot}}$ at insulating boundaries), yields

$$\nabla \cdot \left(\sigma^{(1)}\nabla\varphi\right) = \nabla \cdot \boldsymbol{j}_{oNA} .$$

Using the experimentally determined conductivity $\sigma^{(1)} = 0.8 \times 10^{6}\ \mathrm{S\ m^{-1}}$ (van der Pauw measurement) and substituting for $\boldsymbol{j}_{\mathrm{oNA}}$, we obtain

$$\nabla \cdot \nabla\varphi = \frac{1}{\sigma^{(1)}} \nabla \cdot \left( \left( \hat{\boldsymbol{e}}_z \times \boldsymbol{N}(\boldsymbol{r}) \right) \sum_k \sigma_{jkk}^{(2)} |E_k(\boldsymbol{r})|^2 \right).$$

In Fig. S2A we show the current-density distribution in the 5 µm-wide symmetric crossbar patterned from the 20 nm thick CuMnAs film when a 40 mA current pulse is applied diagonally from the upper right to the lower left by connecting all four contacts to the pulse source. To reproduce the experimental observation, we choose the typical threshold current density of $\mathrm{j_{th}} = 4.2 \times 10^{7}\ \mathrm{A\ cm^{-2}}$, above which the Néel order is switched and reoriented perpendicular to the local current lines. The dotted circles at the two opposite inner corners indicate the regions where the current density exceeds the switching threshold. Fig. S2B shows an example of the electric-potential distribution $\varphi$ when the MIR-illuminated s-SNOM tip (10 µm wavelength) is positioned 3.8 nm above the lower-right corner, over an area where the preceding current pulse has switched the magnetic order. The potential differences at the outer regions of the crossbar, marked by four × symbols, correspond to the measured oNA-voltage according to $V_{oNA} = (V_{A^+} - V_{A^-}) - (V_{B^+} - V_{B^-})$.

Tapping-mode demodulation

To enable direct comparison with the experimentally demodulated signal, the near field was computed for a series of tip–sample separations corresponding to different phases of the tapping oscillation of the s-SNOM tip. For each tip height, the resulting local field distribution was used to calculate the corresponding oNA-voltage map. The final simulated signal was then obtained by demodulating these individual oNA-voltage maps at the second harmonic of the tip oscillation, thereby reproducing the experimental lock-in detection scheme. The resulting demodulated oNA-voltage distribution is presented in Fig. 3D of the main text.

Symmetry considerations and decay over night; damage through high current pulsing

The initial domain configuration of the cross structure before any switching sequence discussed in the paper is shown in Fig. S3A. The two measurement directions are indicated by the voltage measurement symbols with the double arrows. As in the main text, an enhanced sensitivity at the inner corners of the cross is visible, reflecting the locally increased current density at these positions. In addition, a background contribution as discussed in the main text is apparent.

Fig. S3B displays the result of the first significant switching pulse along the direction shown by the framed arrow. While two small domains had previously switched at lower current densities, the pulse shown here induces a clear and extended reorientation of the Néel vector at the corners, consistent with the expected symmetry of the switching mechanism. Upon switching in the reversed direction, the domains invert polarity (Fig. S3C).

After a waiting period of approximately 15 hours without further switching (Fig. S3D), the domain structure largely relaxes towards the energetically favorable multidomain ground state. The configuration becomes, for the most part, indistinguishable from the initial state shown in panel (A), although a residual domain structure persists in the lower-left corner.

Fig. S3E shows the effect of subsequent high-current switching attempts. In this case, a more complex and less reversible domain configuration emerges. The altered switching behaviour is likely associated with current-induced modifications of the material, leading to pinning of antiferromagnetic domains. A corresponding change in topography at the inner corners is visible when comparing the zoomed-in insets of panels (D) and (E). Although the domain pinning did not occur simultaneously with the first observable topographic modification, it is plausible that subsequent high-current pulses contribute to domain formation at the damaged corners which subsequently influences the switching behavior.

Next, we present the recorded voltages on the contact pairs orthogonal to the configuration presented in Fig. 2 of the main text, which detect photocurrents longitudinal with respect to the pulse-aligned Néel vector. Fig. S4 represents an extended Figure to Fig. 2, including the longitudinal measurement in the right column of Fig. S4C, F, and I. Note, that the antiferromagnetic domain state in the corners of the cross structure investigated in the main text was not switched prior to the corresponding switching sequence. Along the opposite diagonal, however, switching experiments had already been performed, with domains pinned permanently, as discussed above in Fig. S3, which are therefore visible. Fig. S4C, F, and I, illustrate that within our sensitivity, no longitudinal contribution originating from these current-pulse-affected corner areas is observed; hence, the photocurrent is purely transverse.

Photocurrent model for the optical nl-AHE

The optical nl-AHE generates a local source current density $\boldsymbol{j}_{oNA}(\boldsymbol{r})$ via a second-order nonlinear process, which we write in tensor form as:

$$\boldsymbol{j}_{oNA,a}(\boldsymbol{r}) = \sum_{b,c} \sigma^{(2)}_{abc}\,(\boldsymbol{N}(\boldsymbol{r}))E_{\boldsymbol{b}}^{\,*}(\boldsymbol{r})E_c(\boldsymbol{r})$$

where $\widehat{\boldsymbol{\sigma}}^{(2)}$ is the second-order nonlinear conductivity tensor (set by crystal symmetry and the local Néel-domain orientation $\boldsymbol{N}(\boldsymbol{r})$) and $\boldsymbol{E}(\boldsymbol{r})$ is the local electric field. As discussed in the main text and in Materials and Methods, for the pulse-written high-symmetry in-plane states with Néel-vector orientations along $\langle 110\rangle$ and in our rotationally symmetric near-field geometry, the dominant tensor components inject a photocurrent in the film plane and transverse to the Néel vector. In the device-level modeling, we approximate the leading nonlinear tensor elements as equal: $\sigma^{(2)} = \sigma_{jkk}$ where $k$ denote mutually orthogonal field directions, $j$ being in the direction of the generated electric current. Fig. S5 shows the frequency dependence of various 2nd order conductivity tensor elements evaluated with a common phenomenological broadening of $\hbar/\tau \sim 3$ meV. Assuming that the net photocurrent averages to zero when the tip probes regions with randomly oriented small (sub-resolution) domains, we simplify the source term, considering Néel vector orientation in the highly symmetric direction $\langle 110\rangle$:

$$\boldsymbol{j}_{oNA}(\boldsymbol{r}) = \left(\hat{\boldsymbol{e}}_z \times \boldsymbol{N}(\boldsymbol{r})\right)\sum_{k} \sigma^{(2)}_{jkk}|E_k(\boldsymbol{r})|^2$$

Here $\hat{\boldsymbol{e}}_z \times \boldsymbol{N}(\boldsymbol{r})$ is a local vector pointing along the generated photocurrent direction (perpendicular to the Néel vector). We set $\sigma^{(2)}_{jkk} = 0$ in areas where the domain size is smaller than the near-field spot.

The local near-field distribution $\boldsymbol{E}(\boldsymbol{r})$ inside the CuMnAs layer is taken from the FEM simulations described in Material and Methods.

Visualizing Reversible Néel-Texture Switching in a 1-µm-Wide CuMnAs Crossbar via Optical nl-AHE Imaging

The 5 $\mu m$-wide device employs writing pulses to overwrite a sub-resolution domain texture. In contrast, we applied the same write-read protocol to 1 µm-wide crossbars (Fig. 4) on 20 $nm$-thick CuMnAs. In Fig. S6 we now show the full crossbar structure, unlike Fig. 4 of the main text, and show that the response is primarily governed by the displacement and reshaping of a pre-existing extended domain pattern. After applying diagonal writing pulses, the oNA-voltage maps show a reversible reconfiguration that is largely confined to the central overlap region, where the current density is highest.

Fig. S6A summarizes the write geometry and voltage readout. The net potential drops along the two orthogonal diagonals are obtained from the simultaneously recorded voltages $V_A = V_{A^+} - V_{A^-}$ and $V_B = V_{B^+} - V_{B^-}$, from which we form the combinations $V_A + V_B$ (Fig. S6B) and $V_A - V_B$ (Fig. S6C). The strongest polarity-dependent changes, which occur at the current-pulse-affected corners are highlighted with dashed bold outlines.

<u>Optical nl-AHE imaging of twin-boundary–governed domain structures in $50\ nm$ CuMnAs films</u>

Fig. S7A shows the topography of a device patterned from a 50 nm-thick CuMnAs/GaP film, with Fig. S7B the corresponding optical nl-AHE map. The voltage was measured between the bottom and right terminal of the cross, with the other legs left floating. The photocurrent signal exhibits a characteristic stripe-like geometry consistent with twin-boundary–correlated domain patterns reported previously by XMLD-PEEM measurements (*50*).

The enhanced signal amplitude observed along the twin-boundary lines suggests that these structural defects influence the local voltage response. One plausible origin is the enhanced electrical conductivity reported along twin boundaries which may be up to an order of magnitude larger than in the surrounding bulk (*51*) and which can modify the local electric field distribution and thereby increase the locally generated photocurrent.

Most notably, in the upper left region of the cross a domain wall is observed to extend across a twin-boundary stripe, with clear polarity reversal of the optical nl-AHE voltage when crossing between adjacent domains. Such polarity changes across twin-correlated domains are consistent with earlier reports discussed in the main text and further confirm that the optical nl-AHE provides a time-reversal–odd readout of antiferromagnetic order.

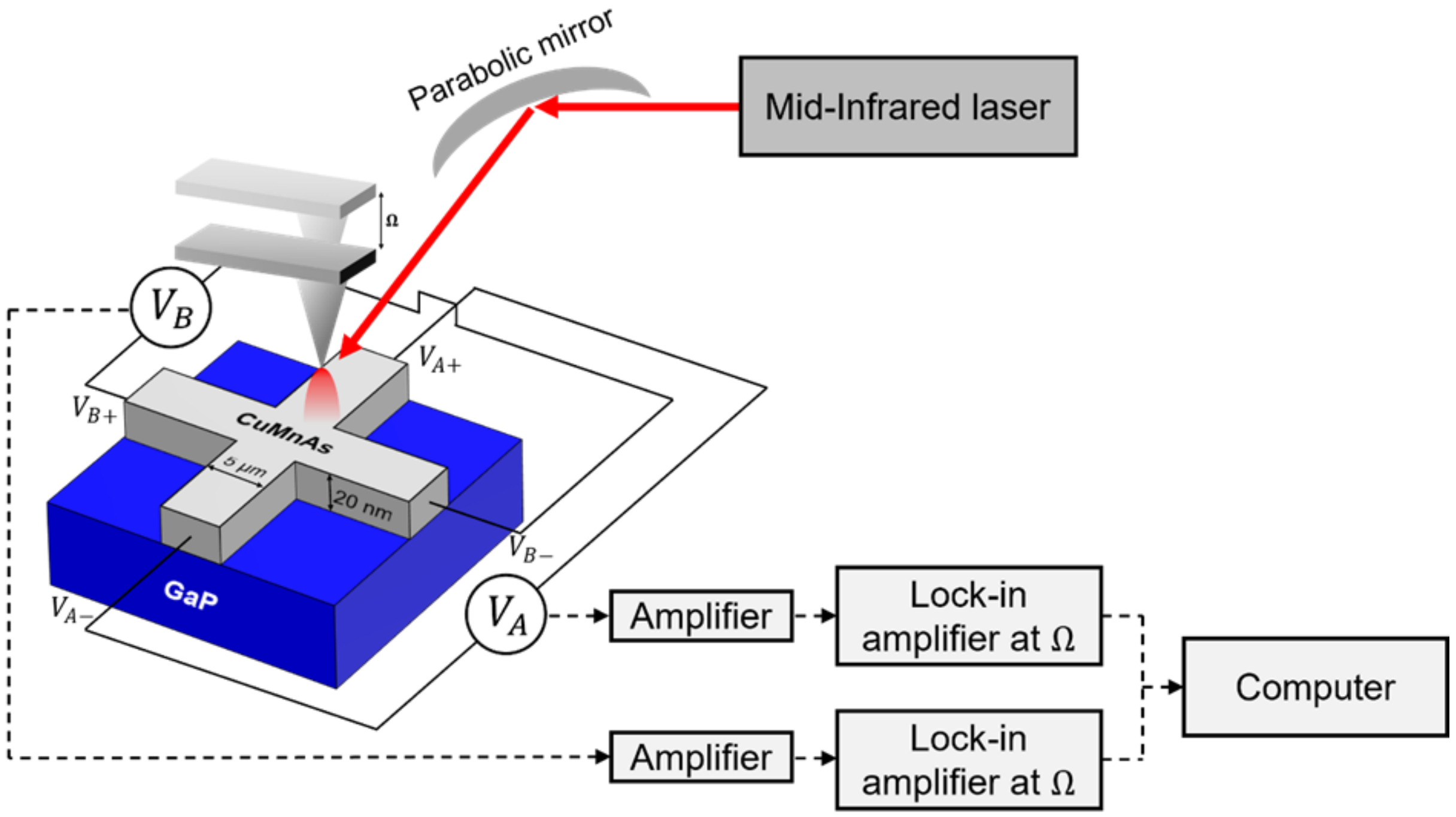

**Fig. S1. Sketch of the photocurrent nanoscopy setup.** A mid-infrared laser is focused on an oscillating s-SNOM tip, where it excites a near-field at the tip apex. This near field drives a photovoltage in the sample that is measured at external contacts along the two arms. After amplification, these voltages are demodulated at higher harmonics of the tip tapping frequency and combined in the post-processing to determine the diagonal voltages.

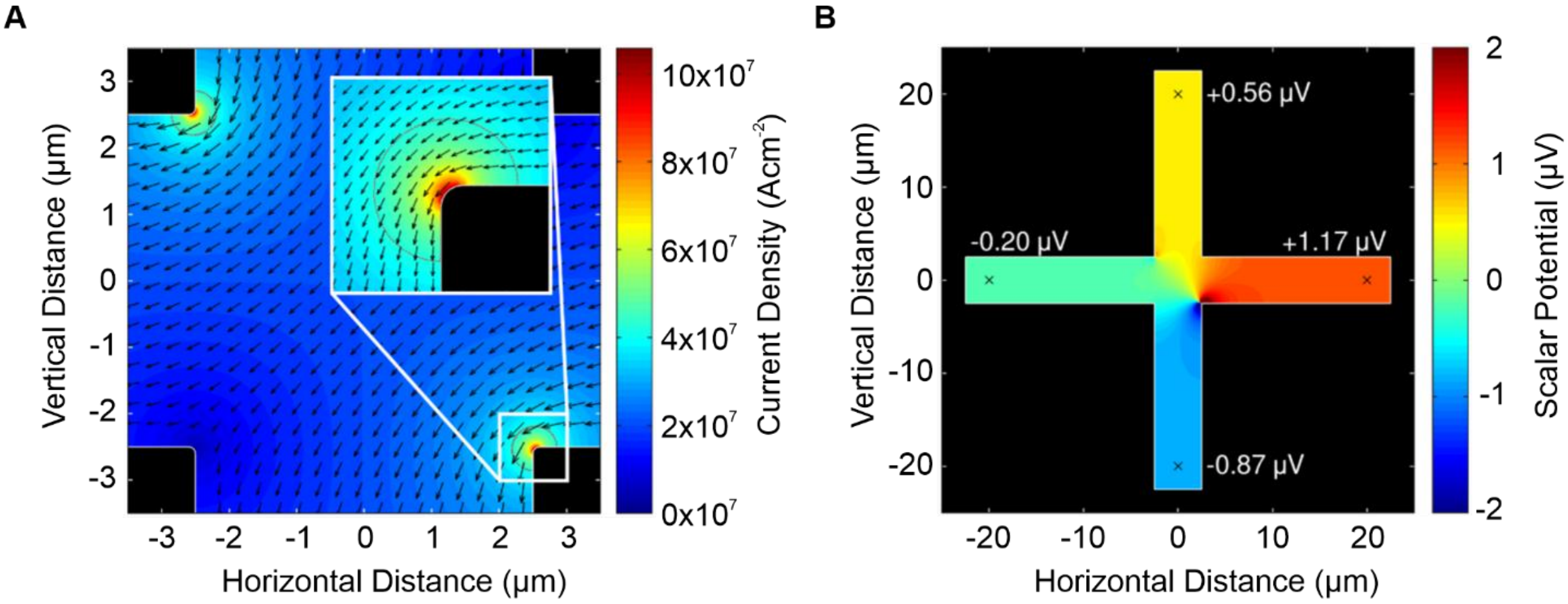

**Fig. S2. Numerical simulation of optical nonlinear anomalous Hall effect in the crossbar geometry.** (**A**) Simulated electric current density magnitude (colour scale) and local direction (arrows) when a 40 mA current pulse is applied to the crossbar. The two circles indicate the inner corners where the current density exceeds the switching threshold value $j_{th} = 4.2 \times 10^7$ A cm$^{-2}$. (**B**) Corresponding steady-state scalar potential φ (colour scale) generated by the optical nl-AHE source current for a representative readout condition with a tip-sample distance of 3.8 nm and with the tip positioned at the lower right corner. Numerical values indicate φ at the positions marked by crosses on the cross arms.

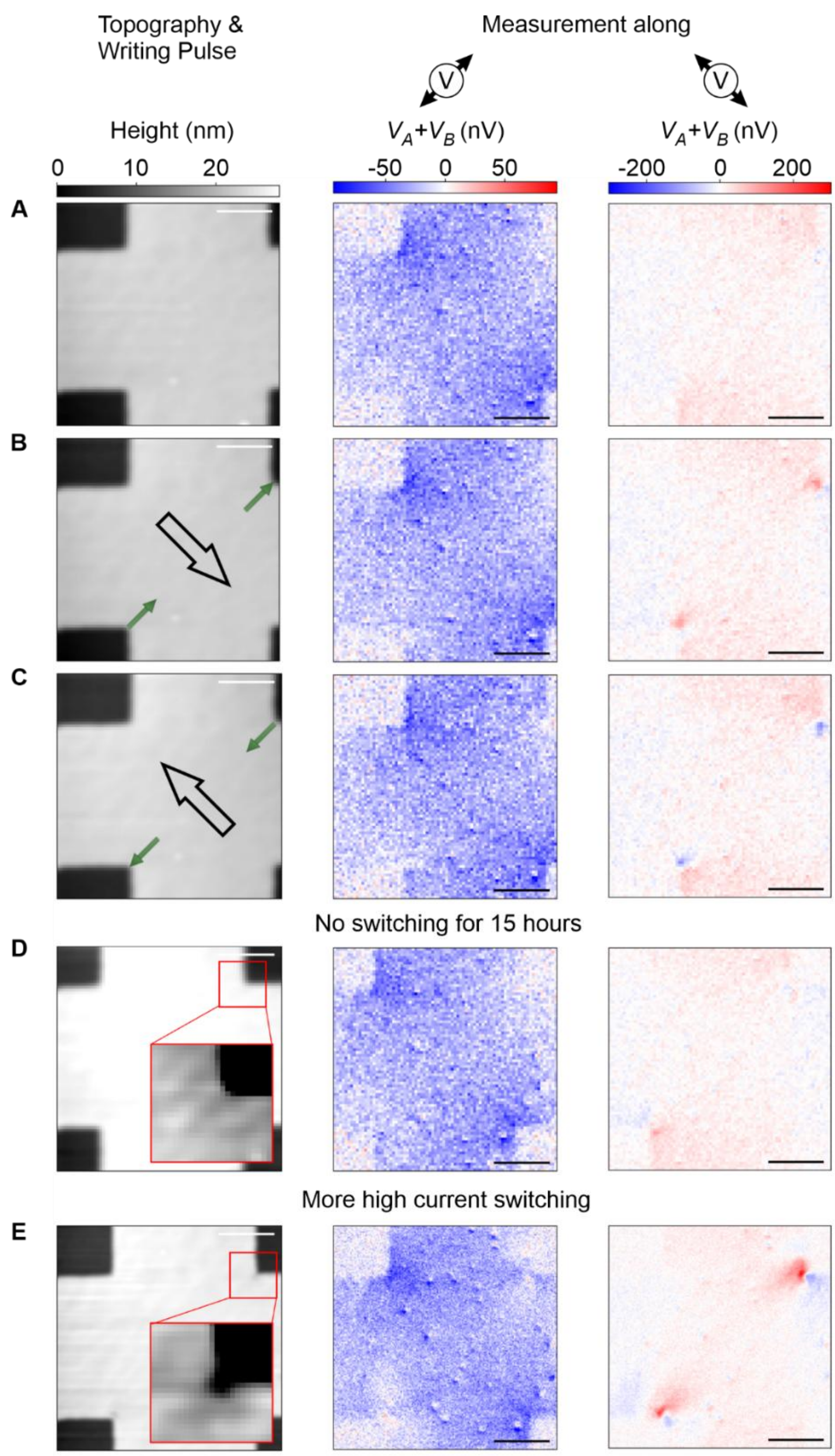

**Fig. S3. Symmetry considerations, decay overnight and damage through high current pulsing.** Shown are the topography and voltage along both diagonals. (**A**) Initial state of the cross, before any switching attempts. (**B**) Switching pulse along the diagonal from the lower left to the upper right, orthogonal to the switching shown in the main paper. (**C**) Inverted switching pulse. (**D**) No switching pulses for 15 hours, allowing the written domains to decay. (**E**) After multiple high current switching pulses, slight damage in the topography (see inset compared to inset in D) appears in the corner and pinned, more complex domain structures develop. The length of the scale bars is 2 µm.

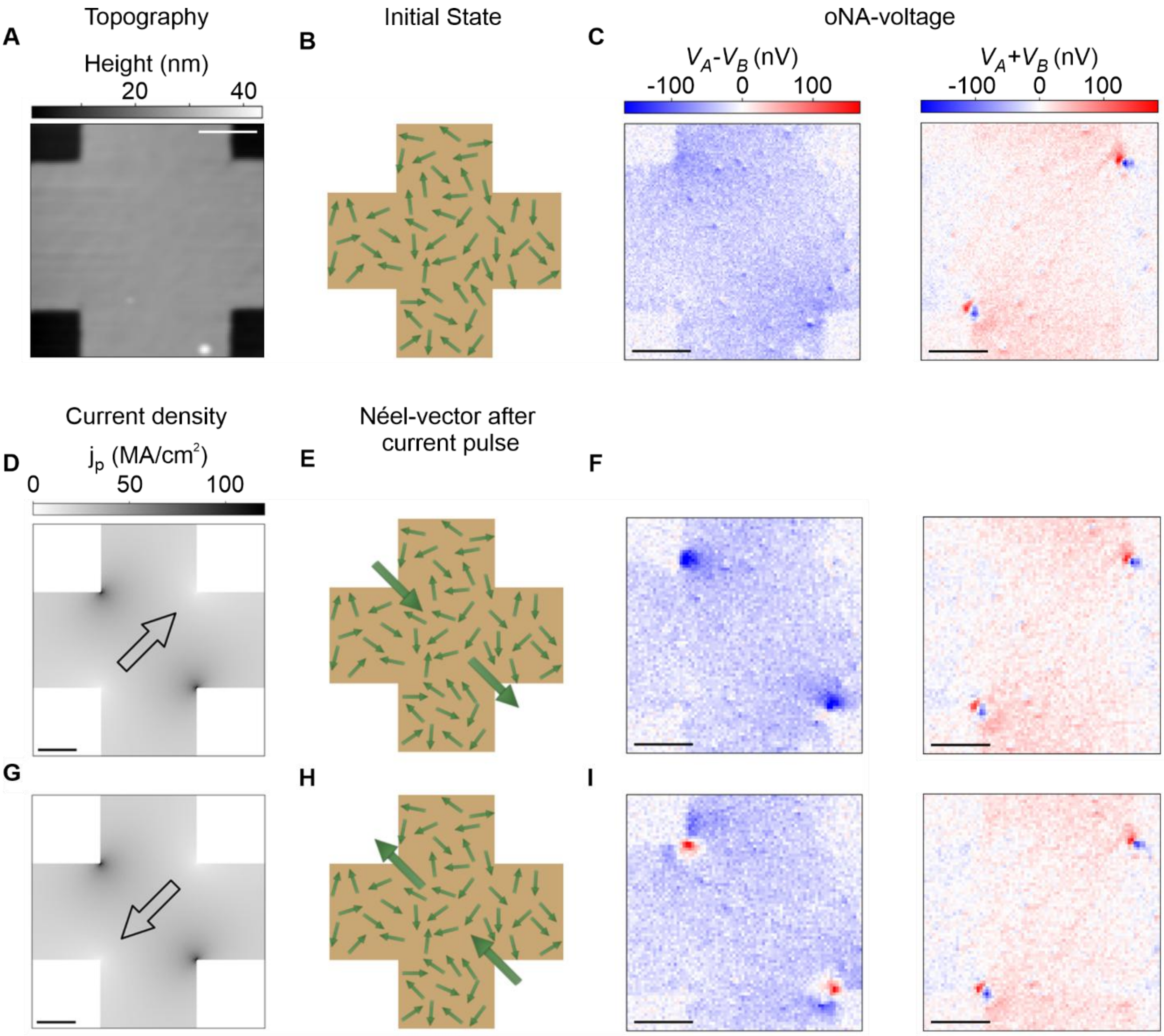

**Fig. S4. Additional data of the switching experiment.** Same as Fig. 2 of the main text, extended by an oNA-voltage map on the right column which was measured along the diagonal between the upper-left and lower-right position.

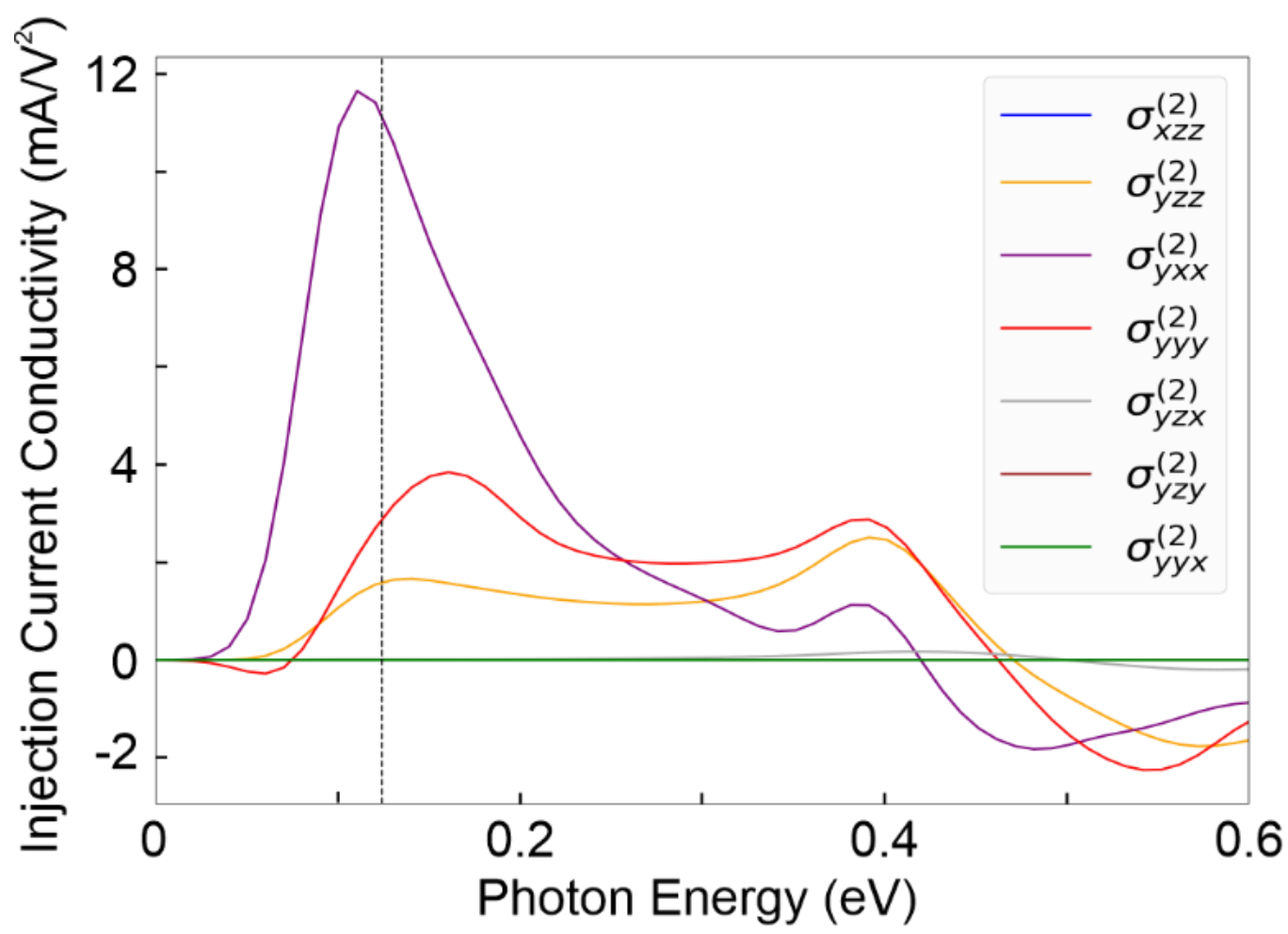

**Fig. S5. Calculated injection-current conductivity components**. Second order conductivity components $\sigma^{(2)}_{\mathrm{abc}}(\omega)$ relevant to the experimental near field s-SNOM geometry with diagonal and off-diagonal terms as indicated are shown, evaluated with a common phenomenological broadening $\hbar/\tau \sim 3$ meV corresponding to $\tau = 235$ fs.

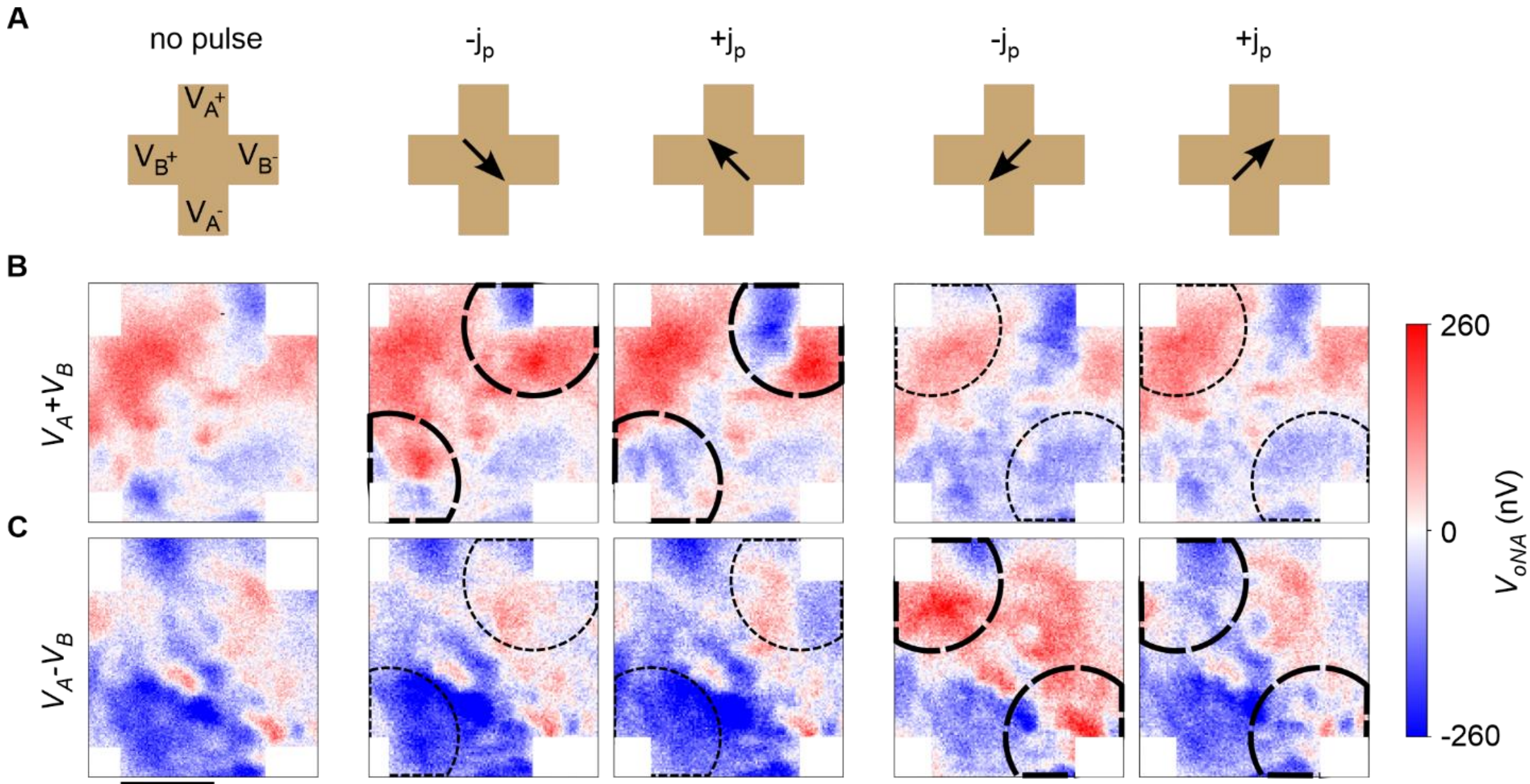

**Fig. S6. Reversible Reconfiguration of Néel Textures in a 1 μm CuMnAs Crossbar Revealed by Optical nl-AHE Imaging.** (**A**) Presets: virgin state and single diagonal writing pulses of opposite polarity along each diagonal (arrows). The simultaneously recorded diagonal voltages are $V_A = V_{A^+} - V_{A^-}$ and $V_B = V_{B^+} - V_{B^-}$. (**B**, **C**) oNA-voltage maps for the presets, shown as $V_A + V_B$ (**B**) and $V_A - V_B$ (**C**). Dashed outlines highlight the inner corners with highest current density and strongest polarity dependence; oNA maps with bold (light) dashed outlines are sensitive to the potential difference along (orthogonal to) the writing-pulse axis. The affected regions swap between channels when the writing diagonal is rotated by 90°, consistent with the same dominant Hall-like current orientation for the high-symmetry ⟨110⟩ states relevant here. (scale bar, 500 nm).

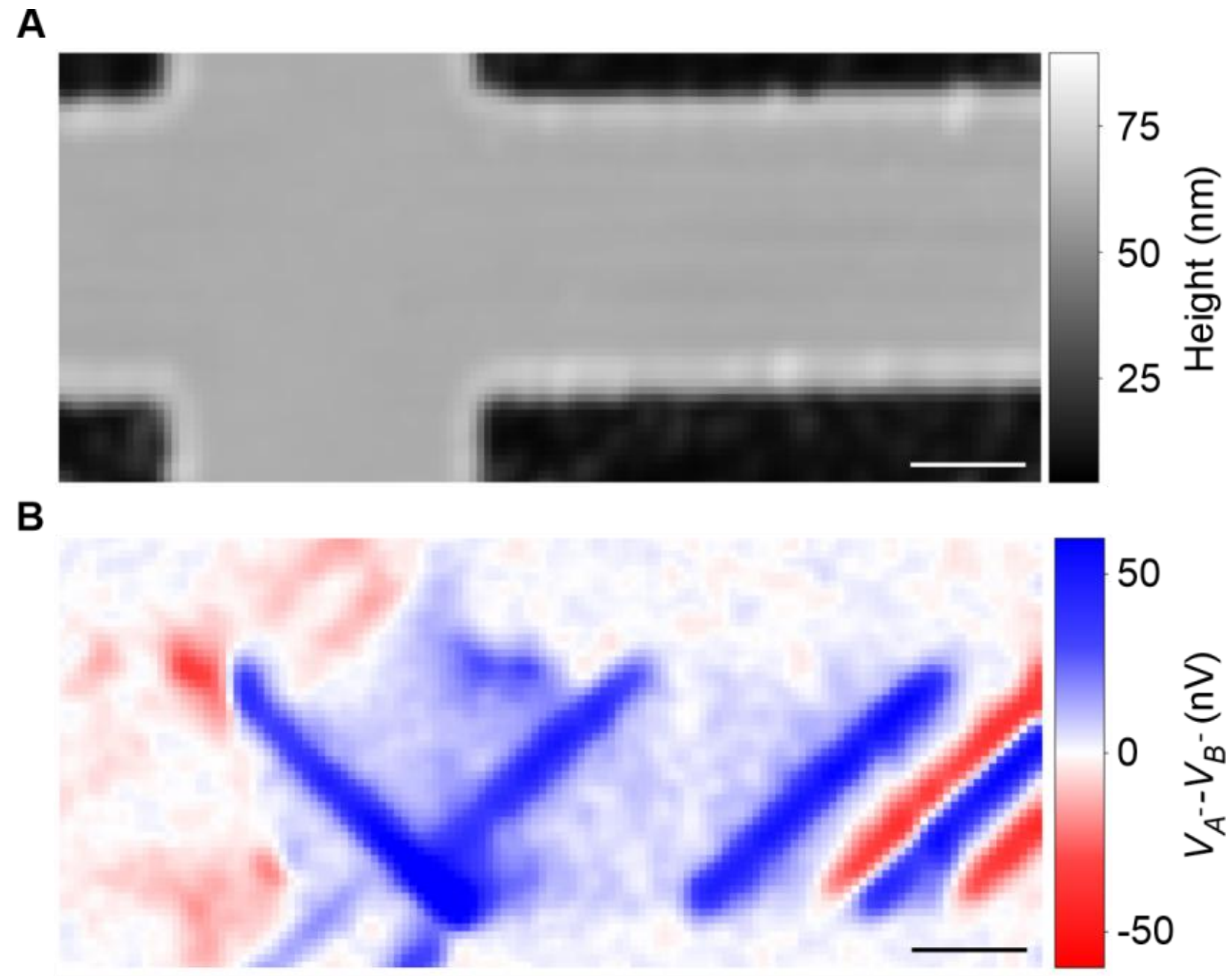

**Fig. S7. Optical nl-AHE imaging of a twin-defect-dominated thick CuMnAs film.** (**A**) AFM topography of a 2 μm-wide crossbar patterned from a 50 nm thick CuMnAs/GaP(001) film (arms along [100] and [010]). (**B**) Corresponding near-field oNA-voltage map acquired at $\lambda = 10$ μm. The stripe-like contrast follows the characteristic defect-correlated morphology of microtwin-defects dominated thick CuMnAs films reported by XMLD-PEEM/XLD-PEEM, with features aligned along the in-plane ⟨110⟩ directions. A representative 180° domain wall shows a sharp sign reversal of the optical nl-AHE contrast across the wall.